# Comment on "Analysis of Scissors Cutting Paper at Super Luminal Speeds"

Chandru Iyer[1]

[1]Plant Head (Retd), Lydall Performance Materials, Gurugram, India

[1]e-mail: chandru_i@yahoo.com

Abstract: In a recent paper, it was demonstrated that the popular legend that scissors can cut paper at super luminal speeds is indeed correct. In this comment we make this additional observation that when the cutting speed is super luminal, the direction of the cutting of the paper as observed by observers in the inertial frame stationary with respect to the blade of the scissors and the direction of the cutting of the paper as observed by observers in the inertial frame stationary with respect to the paper are opposite. This follows from the relativity of simultaneity, and the fact that space-time points that are separated beyond reach at sub-luminal speeds exhibit reversal in time order as observed by different inertial frames. The fact that the paper was cut in opposite directions as observed by the paper and scissors, can be interesting facet for students of Physics.

In a recent paper [1] it was demonstrated that scissors can cut paper at super luminal speeds. While we agree with this, we wish to point out that the direction of the cutting will be reversed as observed by the paper and scissors. We have shown [2] that two inclined rods moving towards each other will collide in such a way that the bottom edge will collide first and the top edge later, as observed by inertial frame co-moving with one rod and vice-versa as observed by inertial frame co-moving with the other rod. Thus the direction of the progression of the collision will be reversed as observed by the observers co-moving with one rod and observers co-moving with the other rod. If we visualize one rod as scissor blade and the other rod as a paper, we can see that the paper will be cut at super luminal speeds and in opposite directions as observed by the two sets of observers co-moving with the scissor blade and the paper respectively.

In a closely related paper [3], examining the rod-slot paradox, we have shown that the leading edge of the rod enters the slot first as observed by the rod, and the trailing edge of the rod enters the slot first as observed by the slot. Here also if the rod can be visualized as a blade and the slot as a sheet of paper we can see that the paper getting cut at super luminal speed but in different directions as observed by the rod (blade) and the slot (paper)

Both the 'collision of inclined rods' and the 'rod and slot' problems have also been analyzed and solved by transformational techniques of spatial and space-time rotations [4] yielding the same results as in [2] and [3]

Conclusion: While it is interesting to note that paper can be cut at super luminal speeds, it is additionally intriguing that the direction of the progression of the cutting is reversed as observed by observers stationary with respect to the paper compared to the direction of the progression of the cutting as observed by observers stationary with respect to the blade,